\documentclass{svjour3}                     % onecolumn (standard format)
\smartqed  % flush right qed marks, e.g. at end of proof
\usepackage{graphicx}
\usepackage{fix-cm}
\usepackage{amsmath}
\catcode`\@=11

\@addtoreset{equation}{section}

\journalname{Journal of Engineering Mathematics}

\begin{document}

\title{Relativistic shear-free fluids with symmetry}


\author{S. Moopanar \and  S. D. Maharaj}


\institute{
S. Moopanar  \and  S. D. Maharaj \at
Astrophysics and Cosmology Research Unit,
 School of Mathematics, Statistics and Computer Science,
 University of KwaZulu-Natal, Private Bag X54001, Durban 4000, South Africa
\\ \\
\email{maharaj@ukzn.ac.za}
}

\date{Received: date / Accepted: date}

\maketitle

\begin{abstract}
We study the complete conformal geometry of  shear-free
spacetimes with spherical symmetry and do not specify the form of the matter
content. The general conformal Killing symmetry is solved and we can explicitly exhibit
the vector. The existence of a conformal symmetry places restrictions on the
model. The conditions on the gravitational potentials
are expressed as a system of integrability conditions.
Timelike sectors and inheriting conformal symmetry vectors, which
map fluid flow lines conformally onto fluid flow lines, are
generated and the integrability conditions are shown to be
satisfied. As an example, a spacetime, which is expanding and accelerating, is identified
which contains a spherically symmetric conformal symmetry.


\end{abstract}

\section{Introduction}

The importance and relevance of spacetime symmetries have been emphasised in the reviews of Choquet--Bruhat et al \cite[Chap. 5]{bruhat}, Stephani et al \cite[Chap. 11]{stephani} and Hall \cite[Chap. 7, Chap. 11]{hall}. Symmetries have been studied extensively as they help to provide a deeper insight into the spacetime geometry. They assist in the modelling of realistic astrophysical and cosmological phenomena in a general relativistic setting by providing new exact solutions to the field equations: the nonlinear Einstein equations for neutral barotropic matter and the coupled Einstein--Maxwell equations in the presence of the electromagnetic field. Spacetime symmetries are crucial for the central role they play in the classification and categorisation of the exact solutions of the Einstein field equations. A major thrust is to establish and clarify links between symmetries, conserved quantities, the structure of manifolds and gravitational interactions.

Of particular interest are the conformal symmetries. They have the geometric property of preserving the structure of the null cone by mapping null geodesics to null geodesics. Conformal symmetries are physically significant as they generate constants of the motion along null geodesics for massless particles. The conformal geometry has been studied in Robertson--Walker spacetimes by Maartens and Maharaj \cite{maartens1} and Keane and Barrett \cite{keane1}. A detailed analysis of conformal vectors has been undertaken by Maartens and Maharaj \cite{maartens2} and Keane and Tupper \cite{keane2} in $pp$--wave spacetimes. Tupper et al \cite{tupper} considered the existence of conformal vectors in null spacetimes. The general conformal equations in plane symmetric static spacetimes was completed by Saifullah and Yazdan \cite{saifullah}. The full conformal structure in spherically symmetric static spacetimes was found by Maartens et al \cite{maartens3}. The conformal geometry of nonstatic spherically symmetric spacetimes was analysed by Moopanar and Maharaj \cite{moopanar} without specifying the form of the matter content. Chrobok and Borzeszkowski \cite{chrobok} modelled irreversible thermodynamic processes close to equilibrium and Bohmer et al \cite{bohmer} modelled wormhole structures with exotic matter. These examples illustrate the applications of conformal motions in cosmology. Mak and Harko \cite{mak} studied charged strange stars with a quark equation of state, Esculpi and Aloma \cite{esculpi} generated anisotropic relativistic charged fluid spheres with a linear barotropic equation of state, Usmani et al \cite{usmani} extended the concept of a Bose--Einstein condensate to gravity to construct gravastars and Herrera et al \cite{herrera1} studied reversible dissipative processes and Landau damping in stellar systems. These examples are a sample of the various applications of conformal motions in relativistic astrophysics. Therefore it is important to find conformal symmetries in spacetimes of physical interest.

Shear--free relativistic fluids which are expanding and accelerating are important for describing gravitational processes in inhomogeneous cosmological models \cite[Chap. 2]{krasinski} and radiating stellar spheres \cite{herrera2}. There is a need to study the conformal symmetries in such  settings. Our intention is to generate the general conformal Killing vector for an arbitrary matter distribution. The dynamics are studied by considering the Einstein field equations in a manifold under the action of the Lie algebra of conformal vectors. In Section 2 the shear--free line element and conformal Killing equations are presented. The system of coupled conformal equations are solved and the general solution is presented explicitly subject to integrability conditions. The particular cases of timelike hypersurface orthogonal and inheriting conformal symmetries are considered in Section 3. The integrability conditions can be integrated and the conformal vector has a simple form. We demonstrate in Section 4 exact solutions of the Einstein field equations are possible for expanding and accelerating relativistic fluids by assuming a particular spherically symmetric conformal vector. Some concluding remarks are made in Section 5.

\section{Conformal Equations}

We write the line element for shear--free spherically symmetric spacetimes in the form
\begin{equation}
\mathrm{d}s^2 = - \mathrm{e}^{2\nu(t,r)} \mathrm{d}t^2 + \mathrm{e}^{2\lambda(t,r)}
\left[ \mathrm{d}r^2 + r^2(\mathrm{d}\theta^2 + \sin^2 \theta \mathrm{d}\phi^2) \right],
\label{sfss}
\end{equation}
where the gravitational potentials $\nu$ and $\lambda$ are functions of $t$ and $r$. We are utilising coordinates $(x^a) = (t,r,\theta,\phi)$ that are simultaneously comoving and isotropic relative to the fluid four--velocity $\displaystyle u^i = \mathrm{e}^{-\nu} \delta^i_{{}0}$. The nonvanishing kinematical quantities for (\ref{sfss}) are the acceleration and the expansion. As this spacetime is spherically symmetric, the vorticity vanishes in comoving coordinates.

A conformal Killing vector $\bf X$ satisfies the relationship
\begin{equation}
{\cal L}_{\bf X} g_{ab} = 2 \psi g_{ab},      \label{ckv}
\end{equation}
which determines the action of the Lie infinitesimal generator ${\cal L}_{\bf X}$ on the metric tensor field $\bf g$. The quantity $\psi = \psi(x^i)$ is the conformal factor. It is possible to analyse the conformal symmetry structure of the shear--free metric (\ref{sfss}) without having to assume forms for the conformal vector $\bf X$ and the conformal factor $\psi$. The conformal Killing vector equation (\ref{ckv}) generates the following system for the metric (\ref{sfss}):
\begin{subequations}  \label{system}
\begin{eqnarray}
\nu_t X^0 + \nu_r X^1 +
X^0_{{}t} &=& \psi,  \label{system1} \\
\mathrm{e}^{2\lambda} X^1_{{}t} - \mathrm{e}^{2\nu} X^0_{{}r} &=& 0, \label{system2} \\
r^2 \mathrm{e}^{2\lambda} X^2_{{}t} - \mathrm{e}^{2\nu} X^0_{{}\theta} &=& 0,
\label{system3} \\
r^2 \mathrm{e}^{2\lambda}\sin^2 \theta  X^3_{{}t} - \mathrm{e}^{2\nu} X^0_{{}\phi}
&=& 0, \label{system4} \\
\lambda_t X^0 + \lambda_r X^1 +
X^1_{{}r} &=& \psi, \label{system5} \\
r^2 X^2_{{}r} + X^1_{{}\theta} &=& 0, \label{system6} \\
r^2 \sin^2 \theta X^3_{{}r} + X^1_{{}\phi} &=& 0, \label{system7} \\
\lambda_t X^0 + (r^{-1} + \lambda_r) X^1 +
X^2_{{}\theta} &=& \psi,  \label{system8} \\
\sin^2 \theta X^3_{{}\theta} + X^2_{{}\phi} &=& 0, \label{system9} \\
\lambda_t X^0 + (r^{-1} + \lambda_r) X^1 +
\cot \theta X^2 + X^3_{{}\phi} &=& \psi. \label{system10}
\end{eqnarray}
\end{subequations}
This system comprises first order coupled partial differential equations which are linear in the conformal vector ${\bf X} = (X^0,X^1,X^2,X^3)$ and the conformal factor $\psi$. In our notation subscripts denote partial differentiation.

The integration procedure to obtain the conformal Killing vector involves manipulating the above system to obtain equations involving individual components $X^0$, $X^1$, $X^2$, $X^3$ and $\psi$. The resulting equations decouple and integrability conditions (restricting the metric functions $\nu$ and $\lambda$) are generated. We find that the components of the conformal Killing vector and the conformal factor are given by
\begin{eqnarray}
X^0 &=& r^2 \mathrm{e}^{2(\lambda-\nu)} \sin\theta ({\cal C}_t \sin\phi - {\cal
D}_t \cos\phi) - r^2 \mathrm{e}^{2(\lambda-\nu)} {\cal I}_t \cos\theta +
{\cal J}(t,r),   \nonumber \\
X^1 &=& - r^2 \sin\theta ({\cal C}_r \sin\phi - {\cal
D}_r \cos\phi) + r^2 {\cal I}_r \cos\theta +
{\cal K}(t,r),   \nonumber \\
X^2 &=& \cos\theta \left[ {\cal C}(t,r) \sin\phi - {\cal
D}(t,r) \cos\phi \right] +
\cos\theta (a_1 \sin\phi - a_2
\cos\phi) - a_3 \sin\phi \nonumber \\ & & + a_4 \cos\phi
+ {\cal I}(t,r) \sin\theta, \nonumber \\
X^3 &=& \csc\theta \left[ {\cal C}(t,r) \cos\phi + {\cal
D}(t,r) \sin\phi \right] +
\csc\theta (a_1 \cos\phi + a_2
\sin\phi) \nonumber \\ & & - \cot\theta (a_3 \cos\phi + a_4 \sin\phi)
+ a_5, \nonumber \\
\psi &=& r^2 \sin\theta \sin\phi
\left[\mathrm{e}^{2(\lambda-\nu)}{\cal C}_{tt} +
(2\lambda_t - \nu_t ) \mathrm{e}^{2(\lambda-\nu)}{\cal C}_{t} - \nu_r {\cal
C}_r \right] \nonumber \\ & & -
r^2 \sin\theta \cos\phi
\left[\mathrm{e}^{2(\lambda-\nu)}{\cal D}_{tt} +
(2\lambda_t - \nu_t ) \mathrm{e}^{2(\lambda-\nu)}{\cal D}_{t} - \nu_r {\cal
D}_r  \right] \nonumber \\ & & -
r^2 \cos\theta \left[ \mathrm{e}^{2(\lambda-\nu)}{\cal I}_{tt} +
(2\lambda_t -\nu_t) \mathrm{e}^{2(\lambda-\nu)} {\cal I}_t -
\nu_r {\cal I}_r \right] \nonumber \\ & & +
{\cal J}_t +
\nu_t {\cal J}(t,r) +
\nu_r {\cal K}(t,r), \nonumber
\end{eqnarray}
where ${\cal C}$, ${\cal D}$, ${\cal I}$, ${\cal J}$ and ${\cal K}$ are
arbitrary functions of $t$ and $r$ that arise from the integration
process and $a_1$--$a_5$ are constants. This solution is
subject to the following twelve integrability conditions:
\begin{eqnarray}
{\cal C}_{tr} + ( r^{-1} + \lambda_r
- \nu_r ) {\cal C}_{t} &=& 0,  \nonumber \\
{\cal D}_{tr} + ( r^{-1} + \lambda_r
- \nu_r ) {\cal D}_{t} &=& 0,  \nonumber \\
{\cal I}_{tr} + ( r^{-1} + \lambda_r
- \nu_r ) {\cal I}_{t} &=& 0,  \nonumber \\
\mathrm{e}^{2(\lambda - \nu)}{\cal C}_{tt} + {\cal C}_{rr} +
(\lambda_{t}-\nu_t) \mathrm{e}^{2(\lambda - \nu)}{\cal C}_{t} +
(2r^{-1}+\lambda_{r}-\nu_r) {\cal C}_{r}
&=& 0,  \nonumber \\
\mathrm{e}^{2(\lambda - \nu)}{\cal D}_{tt} + {\cal D}_{rr} +
(\lambda_{t}-\nu_t) \mathrm{e}^{2(\lambda - \nu)}{\cal D}_{t} +
(2r^{-1}+\lambda_{r}-\nu_r) {\cal D}_{r}
&=& 0,   \nonumber \\
\mathrm{e}^{2(\lambda-\nu)} {\cal I}_{tt} +
{\cal I}_{rr} +
(\lambda_t - \nu_t ) \mathrm{e}^{2(\lambda-\nu)} {\cal I}_{t} +
(2 r^{-1} + \lambda_{r} - \nu_r ) {\cal I}_{r}
&=& 0,  \nonumber \\
r^2 \mathrm{e}^{2(\lambda-\nu)}{\cal C}_{tt} + r^2 (\lambda_t - \nu_t )
\mathrm{e}^{2(\lambda-\nu)}{\cal C}_{t} + r^2
(r^{-1} + \lambda_{r} - \nu_r ) {\cal C}_{r} \,+\, {\cal C} + a_1
&=& 0,   \nonumber \\
r^2 \mathrm{e}^{2(\lambda-\nu)}{\cal D}_{tt} + r^2 (\lambda_t - \nu_t )
\mathrm{e}^{2(\lambda-\nu)}{\cal D}_{t} + r^2
(r^{-1} + \lambda_{r} - \nu_r ) {\cal D}_{r} + {\cal D} + a_2
&=& 0,   \nonumber \\
r^2 \mathrm{e}^{2(\lambda-\nu)} {\cal I}_{tt} +
r^2 (\lambda_t - \nu_t) \mathrm{e}^{2(\lambda-\nu)} {\cal I}_{t} +
r^2 (r^{-1} + \lambda_{r} - \nu_r ) {\cal I}_{r} + {\cal I}
&=& 0,  \nonumber \\
\mathrm{e}^{2\lambda} {\cal K}_{t} - \mathrm{e}^{2\nu} {\cal J}_{r} &=& 0,
\nonumber \\
- {\cal J}_t +
(\lambda_t - \nu_t ){\cal J} +
(r^{-1} + \lambda_{r} - \nu_r ) {\cal K}
&=& 0,  \nonumber \\
- {\cal J}_t +
{\cal K}_r +
(\lambda_t - \nu_t ) {\cal J} +
(\lambda_r - \nu_r ) {\cal K}
&=& 0.  \nonumber
\end{eqnarray}
These conditions place restrictions on the functions of integration and
the metric functions.

The above solution may be simplified and
expressed in a more
compact form. The constants $a_1$ and $a_2$ are eliminated if we let
$$
\tilde{\cal C} = {\cal C} + a_1 \quad \mbox{and}
\quad \tilde{\cal D} = {\cal D} + a_2.
$$
We introduce the new variables
\begin{eqnarray}
A^i &=& (A^1,A^2,A^3) = \left( \tilde{\cal C}, -\tilde{\cal D}, -{\cal I} \right),  \nonumber \\
\eta_i &=& (\eta_1,\eta_2,\eta_3) = \left( \sin\theta \sin\phi, \sin\theta \cos\phi, \cos\theta
\right), \nonumber \\
A^0 &=& {\cal J}(t,r), \nonumber \\
A^4 &=& {\cal K}(t,r).  \nonumber
\end{eqnarray}
The components of the conformal Killing vector and the conformal factor can then be written in
the form
\begin{subequations} \label{soln}
\begin{eqnarray}
X^0 &=& r^2 \mathrm{e}^{2(\lambda-\nu)} A^i_{{}t} \eta_i + A^0, \label{soln1} \\
X^1 &=& -r^2 A^i_{{}r} \eta_i + A^4, \label{soln2} \\
X^2 &=& A^i (\eta_i)_{{}_\theta} - a_3\sin\phi + a_4\cos\phi,
\label{soln3} \\
X^3 &=& \csc^2\theta A^i (\eta_i)_{{}_\phi} - \cot\theta
(a_3\cos\phi+a_4\sin\phi) + a_5,  \label{soln4} \\
\psi &=& r^2 \eta_i \left[ e^{2(\lambda-\nu)} A^i_{{}tt} +
(2\lambda_t-\nu_t) \mathrm{e}^{2(\lambda-\nu)}
A^i_{{}t} - \nu_r A^i_{{}r} \right]  \nonumber \\
&& + A^0_{{}t}
+ \nu_t A^0
 + \nu_r A^4.  \label{soln5}
\end{eqnarray}
\end{subequations}
The integrability conditions can also be written more compactly as
\begin{subequations}  \label{ic}
\begin{eqnarray}
A^i_{{}tr} + (r^{-1}+\lambda_r-\nu_r) A^i_{{}t} &=& 0,  \label{ic1}
\\
\mathrm{e}^{2(\lambda-\nu)} A^i_{{}tt} + A^i_{{}rr} + (\lambda_t-\nu_t)
\mathrm{e}^{2(\lambda-\nu)} A^i_{{}t}  && \nonumber \\
 + (2r^{-1}+\lambda_r-\nu_r) A^i_{{}r}
&=& 0,  \label{ic2} \\
r^2 \mathrm{e}^{2(\lambda-\nu)} A^i_{{}tt} +
r^2 (\lambda_t-\nu_t)
\mathrm{e}^{2(\lambda-\nu)} A^i_{{}t}  \nonumber \\
+ r^2(r^{-1}+\lambda_r-\nu_r)
A^i_{{}r}  + A^i &=& 0,  \label{ic3} \\
\mathrm{e}^{2\lambda} A^4_{{}t} - \mathrm{e}^{2\nu} A^0_{{}r} &=& 0,  \label{ic4} \\
-A^0_{{}t} + (\lambda_t-\nu_t)A^0 + (r^{-1}+\lambda_r-\nu_r)A^4
&=& 0,  \label{ic5} \\
-A^0_{{}t} + A^4_{{}r}
+ (\lambda_t-\nu_t)A^0 + (\lambda_r-\nu_r)A^4
&=& 0.  \label{ic6}
\end{eqnarray}
\end{subequations}
The conformal vector $\bf X$ and the conformal factor $\psi$ in
(\ref{soln}) constitute the general solution of the system (\ref{system}).
This solution is subject to the integrability conditions (\ref{ic}). The
angular dependence of the solution is fully determined as the spacetime
is spherically symmetric; the functions of integration and the metric
potentials depend only on $t$ and $r$. The general conformal symmetry found represents a shear--free relativistic fluid which is expanding and accelerating under the action of (\ref{ckv}).

It is possible to achieve a simpler form of the conformal vector $\bf X$ in the shear--free model studied here than is possible in the shearing model of \cite{moopanar} since there are fewer metric functions. We achieve this further simplification by taking various combinations of the conditions (\ref{ic}) to obtain the conditions
\begin{eqnarray}
r^2 A^i_{rr} + r A^i_r - A^i &=& 0, \nonumber \\
A^4_{{}r} - r^{-1} A^4 &=& 0. \nonumber
\end{eqnarray}
Upon integration we get
\begin{eqnarray}
A^i &=& r F^i(t) + r^{-1} G^i(t), \nonumber \\
A^4 &=& r F^4(t), \nonumber
\end{eqnarray}
where $F^i$, $G^i$ and $F^4$ are functions of integration. The integrability conditions then imply
$$ \left( r^2 \dot{F}^i + \dot{G}^i \right)_r + (\lambda_r -
\nu_r) \left( r^2 \dot{F}^i + \dot{G}^i \right) = 0,
$$
where dots denote derivatives with respect to $t$. This gives
$$
\mathrm{e}^{(\lambda-\nu)} \left( r^2 \dot{F}^i + \dot{G}^i \right)
= H^i(t),
$$
where the quantities  $H^i$ result from the integration
process. Then with these results we can write (\ref{soln}) in the simpler form
\begin{subequations} \label{solnnew}
\begin{eqnarray}
X^0 &=& r \mathrm{e}^{(\lambda-\nu)} \eta_i H^i + A^0,
\label{solnnew1} \\
X^1 &=& -r^2 \eta_i \left( {F}^i -
r^{-2} {G}^i \right)
+ r F^4,  \label{solnnew2} \\
X^2 &=& (\eta_i)_{{}_\theta}  \left( r F^i +
r^{-1} G^i \right)
- a_3 \sin\phi + a_4 \cos\phi,
\label{solnnew3} \\
X^3 &=& \csc^2\theta (\eta_i)_{{}_\phi} \left( r F^i +
r^{-1} G^i \right)
- \cot\theta (a_3 \cos\phi + a_4 \sin\phi) + a_5,
\label{solnnew4} \\
\psi &=& r \eta_i \left[ \mathrm{e}^{(\lambda-\nu)} \dot{H}^i
+ \lambda_t
\mathrm{e}^{(\lambda-\nu)} {H}^i
- r
\nu_r \left( F^i - r^{-2}G^i \right) \right] \nonumber \\
&& + A^0_t + \nu_t A^0
 + r \nu_r F^4.
\label{solnnew5}
\end{eqnarray}
\end{subequations}
The integrability conditions (\ref{ic}) become
\begin{subequations}  \label{icnew}
\begin{eqnarray}
\mathrm{e}^{(\lambda-\nu)} \left( r^2 \dot{F}^i + \dot{G}^i \right)
&=& H^i, \label{icnew1} \\
\mathrm{e}^{(\lambda-\nu)} \dot{H}^i
+ r(\lambda_r - \nu_r) \left( {F^i} - r^{-2} {G^i}
\right)  + 2 {F^i} &=& 0,  \label{icnew2} \\
r \mathrm{e}^{2\lambda} \dot{F}^4 - \mathrm{e}^{2\nu} A^0_r &=& 0,
\label{icnew3} \\
-A^0_t + F^4 + (\lambda_t - \nu_t) {A^0} + r
(\lambda_r -\nu_r ) {F^4}
 &=& 0.     \label{icnew4}
\end{eqnarray}
\end{subequations}
Observe that only the integration function $A^0$  depends on $t$ and
$r$. The $r$--dependence of the functions $A^i$ and $A^4$ has been
completely specified -- there is freedom only in the $t$ coordinate.

\section{Particular Vectors}

For a purely timelike vector we must have for consistency $F^i = G^i = H^i = 0$, $F^4 = 0$ and $a_3 = a_4 = a_5 = 0$. Therefore the conformal vector assumes the form ${\bf X} =
(A^0,0,0,0)$ with $A^0 \neq 0$. Now the condition (\ref{icnew3})
implies that $A^0 = A^0(t)$ and (\ref{icnew4}) becomes
$$
\dot{A}^0 + (\nu_t-\lambda_t) A^0 = 0,
$$
which on integration yields $A^0 \propto e^{\lambda-\nu}$. Hence we have found that spacetimes exist  with vanishinging shear that admit a timelike conformal vector parallel to the fluid 4--velocity $\bf u$.

The condition
\begin{equation}
{\cal L}_{\bf X} u_a = \psi u_a,   \label{iv}
\end{equation}
was introduced by Herrera et al \cite{herrera3} and Maartens et al \cite{maartens4} and studied widely by Coley and Tupper \cite{coley1}, \cite{coley2} amongst others. Conformal Killing vectors $\bf X$ satisfying (\ref{iv}) are called inheriting as fluid flow lines are mapped conformally. The inheriting condition (\ref{iv}) for the comoving velocity $\displaystyle u^i = e^{-\nu} \delta^i_{{}0}$ generates the following system
\begin{subequations} \label{ivsystem}
\begin{eqnarray}
\nu_t X^0 + \nu_r X^1 + X^0_{{}t} &=& \psi,  \label{ivsystem1}  \\
X^0_{{}r} &=& 0, \label{ivsystem2}  \\
X^0_{{}\theta} &=& 0, \label{ivsystem3}  \\
X^0_{{}\phi} &=& 0. \label{ivsystem4}
\end{eqnarray}
\end{subequations}
The inheriting conditions may be solved since $X^0 = X^0(t)$ and the integration is simplified. Two cases arise since $A^0 = A^0(t)$ and $H^i = 0$; $F^i$ and $G^i$ are constants. The quantity $F^4$ is also constant. Firstly when $F^i \neq 0$ and $G^i \neq 0$ then the integrability conditions become
\begin{subequations} \label{iciv}
\begin{eqnarray}
r(\lambda_r - \nu_r) \left( F^i - r^{-2} G^i
\right)  + 2 F^i &=& 0,  \label{iciv1} \\
-\dot{A}^0 + F^4 + (\lambda_t - \nu_t) {A^0} + r
(\lambda_r -\nu_r ) {F^4}
 &=& 0,     \label{iciv2}
\end{eqnarray}
\end{subequations}
which places a restriction on the potentials $\nu$ and $\lambda$.
Secondly when $F^i = G^i = 0$ then
\begin{subequations} \label{ivsoln}
\begin{eqnarray}
X^0 &=& A^0, \label{ivsoln1} \\
X^1 &=& r F^4,  \label{ivsoln2} \\
X^2 &=& -a_3 \sin\phi + a_4 \cos\phi, \label{ivsoln3} \\
X^3 &=& -\cot\theta (a_3 \cos\phi + a_4 \sin\phi) + a_5,
\label{ivsoln4} \\
\psi &=& \dot{A}^0 + \nu_t A^0 + r \nu_r F^4,  \label{ivsoln5}
\end{eqnarray}
\end{subequations}
and only (\ref{iciv2}) is applicable. Equation (\ref{iciv2}) can be solved by the method of characteristics. The general solution is given by
\begin{equation}
 \lambda-\nu = F(u) + \ln A^0 - \ln r, \label{icivsoln}
\end{equation}
where $\displaystyle u = \int \frac{\mathrm{d}t}{A^0} - \frac{\ln r}{F^4}$ and $F(u)$ is an arbitrary function. Thus we have found the general inheriting conformal symmetry for the shear--free line element (\ref{sfss}). The metric functions $\nu$ and $\lambda$ are restricted by (\ref{icivsoln}) for such a symmetry to exist. Further restrictions may arise on the application of the Einstein field equations. It is for this reason that spacetimes admitting inheriting conformal Killing vectors are rare.

\section{Spherical Models}

The objective is to find an exact solution to the field equations which contains a conformal vector. In our case
the ideal is to obtain the conformal vector $\bf X$ that satisfies the
integrability conditions (\ref{icnew}) and the Einstein field equations. This is difficult to achieve in general. We can make progress if
we suppose that
\begin{equation}
{\bf X} = t \frac{\partial}{\partial t} + r
\frac{\partial}{\partial r},   \label{sscon}
\end{equation}
i.e. the conformal vector is spherically symmetric. This case was first
investigated by Dyer et al \cite{dyer}. For the conformal vector
(\ref{sscon}) the equations (\ref{solnnew}) imply $F^i = G^i = H^i
= 0$, $a_3= a_4 = a_5 = 0$, $A^0 = t$ and $F^4
= 1$. Then the conformal factor (\ref{solnnew5}) becomes
$$
\psi = t \nu_t + r \nu_r + 1,
$$
and the integrability conditions (\ref{icnew}) reduce to the single
equation
$$
t (\lambda_t-\nu_t) + r (\lambda_r-\nu_r) = 1,
$$
with solution
$$\lambda - \nu = \ln t + c\left(\frac{r}{t}\right),$$
for some constant $c$. For the spherically symmetric conformal vector (\ref{sscon})
the solution of the field equations is then reducible to the integration of a
single, nonlinear third order differential equation given by
$$
\mu^2 T_{\mu\mu\mu} + \mu(2m-1) T_{\mu\mu} + (m^2-2m+2T) T_\mu
= 0,
$$
as established by Dyer et al \cite{dyer}. This nonlinear equation may
be integrated in terms of elementary functions and Painlev\'{e}
transcendents as demonstrated by Maharaj  et al \cite{maharaj} and Havas \cite{havas}. The behaviour of the metric functions, governed by this third order equation,
has been comprehensively
investigated by Govinder et al \cite{govinder} utilising the Painlev\'{e}
and Lie analyses.

\section{Discussion}

In this paper  we investigated conformal symmetries in spherically symmetric spacetimes with vanishing shear.
The conformal Killing vector equation was solved in general subject to integrability conditions. We demonstrated
that timeline and inheriting vectors are admitted. As an example we showed that spherically symmetric
conformal symmetries are permitted that satisfy the Einstein field equations. Now that we have  the full
conformal geometry for shear-free relativistic fluids it becomes possible to comprehensively analyze the
Einstein equations to find exact solutions with this symmetry. This is an object of future research.  We have not specified any
matter distribution to find the general conformal symmetry in this analysis. The particular matter field
chosen and the field equations will place restrictions on the dynamics as indicated by Maharaj and Lortan \cite{lortan}.
On physical grounds  it is necessary to specify the form of the matter field when studying the properties of
symmetries. Many treatments of conformal symmetries include energy momentum tensors
comprising of perfect or imperfect matter tensors. The presence of a nonzero electromagnetic field
is likely to produce new effects which are not present in results containing only neutral matter.

\begin{acknowledgements}
 SDM acknowledges that this work
is based upon research supported by the South African Research Chair
Initiative of the Department of Science and Technology and the
National Research Foundation.
\end{acknowledgements}


\begin{thebibliography}{99}

\bibitem{bruhat} Y. Choquet--Bruhat, C. Dewitt-Morette and M. Dillard--Bleick, \textit{Analysis, Manifolds and Physics}. Amsterdam: North--Holland (1982) 656 pp.

\bibitem{stephani} H. Stephani, D. Kramer, M. A. H. MacCallum, C. Hoenselaers and E. Herlt, \textit{Exact
Solutions of Einstein's Field Equations}. Cambridge: Cambridge University Press (2003) 701 pp.

\bibitem{hall} G. S. Hall, \textit{Symmetries and Curvature Structure in General Relativity}. Singapore: World Scientific (2004) 430 pp.

\bibitem{maartens1} R. Maartens and S. D. Maharaj, Conformal Killing Vectors in Robertson--Walker Spacetimes. \textit{Class. Quantum Grav.} 3 (1986) 1005--1011

\bibitem{keane1} A. J. Keane and R. K. Barrett, The Conformal Group SO(4,2) and Robertson--Walker Spacetimes. \textit{Class. Quantum Grav.} 17 (2000) 201--218

\bibitem{maartens2} R. Maartens and S. D. Maharaj, Conformal Symmetries of $pp$--waves. \textit{Class. Quantum Grav.} 8 (1991) 503--514

\bibitem{keane2} A. J. Keane and B. O. J. Tupper, Conformal Symmetry Classes for $pp$--wave Spacetimes. \textit{Class. Quantum Grav.} 21 (2004) 2037--2064

\bibitem{tupper} B. O. J. Tupper, A. J. Keane, G. S. Hall, A. A. Coley and J. Carot, Conformal Symmetry Inheritance in Null Fluid Spacetimes. \textit{Class. Quantum Grav.} 20 (2003) 801--811

\bibitem{saifullah} K. Saifullah and S. Yazdan, Conformal Motions in Plane Symmetric Static Spacetimes. \textit{Int. J. Mod. Phys. D}. 18 (2009) 71--81

\bibitem{maartens3} R. Maartens, S. D. Maharaj and B. O. J. Tupper, General Solution and Classification of Conformal Motions in Static Spherical Spacetimes. \textit{Class. Quantum Grav.} 12 (1995) 2577--2586

\bibitem{moopanar} S. Moopanar and S. D. Maharaj, Conformal Symmetries of Spherical Spacetimes. \textit{Int. J. Theor. Phys.} 49 (2010) 1878--1885

\bibitem{chrobok} T. Chrobok and H. H. Borzeszkowski, Thermodynamical Equilibrium and Spacetime Geometry. \textit{Gen. Relativ. Gravit.} 38 (2006) 397--415

\bibitem{bohmer} C. G. Bohmer, T. Harko and F. S. N. Lobo, Wormhole Geometries with Conformal Motions. \textit{Phys. Rev. D}. 76 (2007) 084014

\bibitem{mak} M. K. Mak and T. Harko, Quark Stars Admitting a One--parameter Group of Conformal Motions. \textit{Int. J. Mod. Phys. D}. 13 (2004) 149--156

\bibitem{esculpi} M. Esculpi and E. Aloma, Conformal Anisotropic Relativistic Charged Fluid Spheres with a Linear Equation of State. \textit{Eur. Phys. J. C}. 67 (2010) 521--532

\bibitem{usmani} A. A. Usmani, F. Rahaman, S. Ray, K. K. Nandi, P. K. F. Kuhfittig, S. A. Rakib and Z. Hasan, Charged Gravastars Admitting Conformal Motion. \textit{Phys. Lett. B}. 701 (2011) 388--392

\bibitem{herrera1} L. Herrera, A. Di Prisco and J. Ibanez, Reversible Dissipative Processes, Conformal Motions and Landau Damping. \textit{Phys. Lett.A}. 376 (2012) 899--900

\bibitem{krasinski} A. Krasinski, \textit{Inhomogeneous Cosmological Models}. Cambridge: Cambridge University Press (1997) 317 pp.

\bibitem{herrera2} L. Herrera and N. O. Santos, Collapsing Spheres Satisfying an "Euclidean Condition". \textit{Gen. Relativ. Gravit.} 42 (2010) 2383--2391

\bibitem{herrera3} L. Herrera, J. Jimenez, L. Leal, J. Ponce de Leon, M. Esculpi and V. Galina, Anisotropic Fluids and Conformal Motions in General Relativity. \textit{J. Math. Phys.} 25 (1984) 3274--3278

\bibitem{maartens4} R. Maartens, D. P. Mason and D. P. Tsamparlis, Kinematic and Dynamic Properties of Conformal Killing Vectors in Anisotropic Fluids. \textit{J. Math. Phys.} 27 (1986) 2987--2994

\bibitem{coley1} A. A. Coley and B. O. J. Tupper, Spherically Symmetric Spacetimes Admitting Inheriting Conformal Killing Vector Fields. \textit{Class. Quantum Grav.} 7 (1990) 2195--2214.

\bibitem{coley2} A. A. Coley and B. O. J. Tupper, Spherical Symmetric Anisotropic Fluid ICKV Spacetimes. \textit{Class. Quantum Grav.} 11 (1994) 2553--2574

\bibitem{dyer} C. C. Dyer, G. C. McVittie and L.M. Oates, A Class of Spherically Symmetric Solutions with Conformal Killing Vectors. \textit{Gen. Relativ. Gravit.} 19 (1987) 887--898

\bibitem{maharaj} S. D. Maharaj, P. G. L. Leach and R. Maartens, Shear--free Spherically Symmetric Solutions with Conformal Symmetry. \textit{Gen. Relativ. Gravit.} 23 (1991) 261--267

\bibitem{havas} P. Havas, Shear--free Spherically Symmetric Perfect Fluid with Conformal Symmetry. \textit{Gen. Relativ. Gravit.} 24 (1992) 599--615

\bibitem{govinder} K. S. Govinder, P. G. L. Leach and S. D. Maharaj, Integrability Analysis of a Conformal Equation in Relativity. \textit{Int. J. Theor. Phys.} 34 (1995) 625--639

\bibitem{lortan}   S. D. Maharaj and D. B. Lortan, Fluid flows with symmetries.
 \textit{Pramana - J. Phys.}
77 (2011) 477--482
\end{thebibliography}
\end{document}